\documentclass[preprint,showpacs]{revtex4}
\usepackage{mathrsfs}
\usepackage{amsfonts}
\usepackage{epsfig}
\usepackage{hyperref}
\usepackage[dvips,usenames]{color}
\usepackage{graphicx}
\usepackage{amsmath}
\begin{document}
\title{Perturbative and nonperturbative contributions to the strange quark asymmetry in the nucleon}

\author{Guan-Qiu Feng $^{1,2}$\footnote{fenggq@ihep.ac.cn},
Fu-Guang Cao $^{3}$\footnote{f.g.cao@massey.ac.nz}, Xin-Heng Guo
$^4$\footnote{xhguo@bnu.edu.cn}, A. I. Signal $^3$\footnote{a.i.signal@massey.ac.nz} } 
\affiliation{$^1$Institute of High Energy Physics, CAS, P.O. Box 918(4), Beijing 100049, China\\
$^2$ Theoretical Physics Center for Science Facilities, CAS, Beijing 100049, China\\
$^3$Institute of Fundamental Science, Massey University, Private Bag 11 222, Palmerston North, New Zealand\\
$^4$College of Nuclear Science and Technology, Beijing Normal
University, Beijing 100875, China }

\begin{abstract}
There are two mechanisms for the generation of an asymmetry between the strange and anti-strange quark distributions in the nucleon:
nonperturbative contributions originating from nucleons fluctuating into virtual baryon-meson pairs such as $\Lambda K$ and $\Sigma K$,
and perturbative contributions arising from gluons splitting into strange and anti-strange quark pairs.
While the nonperturbative contributions are dominant in the large-$x$ region, the perturbative contributions are more significant in the small-$x$ region.
We calculate this asymmetry taking into account both nonperturbative and perturbative contributions, thus giving a more accurate evaluation
of this asymmetry over the whole domain of $x$.
We find that the perturbative contributions are generally a few times larger in magnitude than
the nonperturbative contributions, which suggests that the best region to detect this asymmetry experimentally is in the region $0.02 < x < 0.03$.
We find that the asymmetry may have more than one node,
which is an effect that should be taken into account, {\it e.g.} for parameterizations of the strange and anti-strange quark distributions used in
global analysis of parton distributions.
\end{abstract}
\maketitle

\section{Introduction}
Partons in the sea of the nucleon play an important role in understanding many properties of the nucleon and in explaining many
experimental results in hadronic physics.
For example, it is now well established \cite{PDF_MRST00,PDF_CTEQ5,PDF_GRV98,PDF_Alekhin01}
that gluons carry about 1/2 of the nucleon momentum, while the valence and sea quarks carry the other half.
It is also well known that about 1/3 of the nucleon spin can be attributed to the spin of the valence quarks,
while the rest should be attributed to the spin and orbital angular momenta of quarks and gluons \cite{KuhnCL09,BurkardtMN10}.
Also, subprocesses involving strange and charm quarks of the nucleon contribute $20\%$ to W production at the LHC~\cite{Kusina:2012vh}.
A precise understanding of the strange and charm content of the nucleon is important to the search for physics beyond the Standard Model.

The asymmetry between the strange and anti-strange quarks of the nucleon ($s$-$\bar s$ asymmetry) \cite{Signal:1987gz} could affect the extraction of the Weinberg
angle from neutrino-nucleon deep inelastic scattering \cite{Davidson:2001ji}.
Most theoretical estimations for this asymmetry using quark models of the nucleon suggest (see {\sl e.g.} \cite{Cao:2003ny})
that this effect alone would not be large enough to explain the discrepancy between the NuTeV measurement of ${\rm sin}^2 \theta_{\rm W}$ \cite{Zeller:2001hh} and
the current world-average value \cite{PDG2012}.
Global analysis of hard-scattering data generally provides weak constraints on
the $s$-$\bar s$ asymmetry of the nucleon \cite{Martin:2009iq,Lai:2007dq,Lai:2010,JR09,ABKM10,ABM11,BaronePZ00,Barone:2006xj,NNPDF1.2,NNPDF2.1,NNPDF2.3},
mainly due to the fact that there is limited data sensitive to the strange content of the nucleon.
Some groups have assumed symmetric strange distributions, {\it i.e.} $s=\bar s$, in their latest global fits for the parton distribution functions (PDFs) of the nucleon \cite{Lai:2010,JR09,ABKM10,ABM11},
while other groups were able to  find a strangeness asymmetry that has the magnitude large enough to remove  the NuTeV anomaly
 (see {\it e.g.} \cite{NNPDF1.2,NNPDF2.1,NNPDF2.3,J2010,AKP09}).
In determining the strangeness asymmetry,  most groups adopted functions with one node as the nonperturbative input for the asymmetry
at the initial scale \cite{Martin:2009iq,Lai:2007dq,BaronePZ00,Barone:2006xj,J2010}, although there is no physical principles preventing the asymmetry having multiple nodes.
On this aspect, the neural network technique employed by the NNPDF group has the advantage of allowing for multiple nodes \cite{NNPDF1.2,NNPDF2.1,NNPDF2.3}.

We can classify sea partons according to the dynamics responsible for their creation:
perturbative contributions originating from gluons splitting into quark-antiquark pairs, and nonperturbative contributions,
which come from nucleons fluctuating into baryon-meson pairs with the partons in the baryon-meson pairs manifesting as the sea partons of the nucleon.
The perturbative contributions can be investigated using QCD evolution equations for the parton distribution functions (PDFs) of the nucleon;
the nonperturbative contributions can be estimated using nonperturbative models for the nucleon.
At leading order and next to leading order of the strong coupling constant $\alpha_s$, the perturbative contributions usually respect symmetries of quark models of the nucleon,
such as flavor symmetry, quark-antiquark symmetry and charge symmetry, whereas the nonperturbative contributions generally
break these symmetries \cite{Cao:2011qh}.

The breaking of the symmetry between the strange and anti-strange quark distributions of the nucleon was predicted over two decades
ago by Signal and Thomas \cite{Signal:1987gz} using the meson cloud model (MCM) of the nucleon \cite{Thomas:1983fh}.
The asymmetry they investigated has the nature of a nonperturbative contribution.
Many theoretical papers have since investigated these nonperturbative contributions
\cite{Signal:1987gz,Burkardt:1991di,Brodsky:1996hc,Holtmann:1996be,Cao:1999da,
Melnitchouk:1999mv,Signal:2000xy,Cao:2003ny,Christiansen:1998dz,Cao:2004pr,Alwall:2004rd,
Wakamatsu:2004pd,Alwall:2005xd,Wei:2007nb,Diehl:2007uc,Traini:2011tc}.
These calculations generally predict the distribution $x(s- \bar s)$ peaks in the region of $0.1 < x < 0.2$.

Although it has been generally assumed that perturbatively generated sea distributions respect the symmetries of quark models of the nucleon,
it was shown recently \cite{Catani:2004nc} that the quark-antiquark symmetry of the sea is violated when the perturbative QCD evolution is calculated at
next-to-next-to-leading order (NNLO).
This occurs because the splitting functions for quarks and antiquarks are different from each other at that order \cite{Moch:2004pa}.
To illustrate this evolution effect, a null symmetry at the initial scale, {\it i.e.} $(s-\bar s)(x,Q_0^2)=0$, was assumed in \cite{Catani:2004nc}.
The calculated distribution $x(s -\bar s)$ shows a maximum at $x<0.1$.

In this work we calculate the strange and anti-strange asymmetry in
the whole $x$ region, including both perturbative and
nonperturbative contributions to the asymmetry.
We briefly summarize the formalism for calculating the perturbative and nonperturbative
contributions to the asymmetry in Sec. 2.
The numerical results and discussion, including possible implications for the NuTeV anomaly,
are presented in Sec. 3.
The last section is reserved for a summary.

\section{\label{formalism}Formalism}

\subsection{Nonperturbative Contributions}

In the meson cloud model the nucleon can be viewed as a baryon `core'
surrounded by a mesonic cloud.
The wave function of the nucleon can be expanded in terms of  a series of baryon plus meson Fock
states \cite{Thomas:1983fh},
\begin{eqnarray}\label{nwf}
 N\rangle_{\rm phys} =  Z |N\rangle_{\rm bare} +\sum_{BM}
\sum_{\lambda \lambda^\prime} \int dy \, d^2 {\bf k}_\perp \,
\phi^{\lambda \lambda^\prime}_{BM}(y,k_\perp^2)  |B M \rangle,
\end{eqnarray}
where $Z$ is the wave function renormalization constant,
$\phi^{\lambda \lambda^\prime}_{BM}(y,k_\perp^2)$ is the wave
function of the Fock state containing a virtual baryon ($B$) with
longitudinal momentum fraction $y$, transverse momentum ${\bf
k}_\perp$, and helicity $\lambda$, and a virtual meson ($M$) with
momentum fraction $1-y$, transverse momentum $-{\bf k}_\perp$, and
helicity $\lambda^\prime$.

The lifetime of the virtual baryon-meson components is generally much longer than the interaction
time in the deep inelastic process, thus the quarks and antiquarks in the baryon and meson
contribute to the parton distributions of the nucleon.
These nonperturbative contributions can be expressed as a convolution of fluctuation functions
with the valence parton distributions in the baryon B or meson M.
The contributions to the asymmetry distribution at the scale $Q$, $(s -\bar s)(x, Q^2), $ can be written as,
\begin{eqnarray}
(s -\bar s)  (x, Q^2)
&=&  \int^1_x \frac{dy} {y} \left[ f_{BM/N} (y) s_B(\frac{x}{y}, Q^2) \right. \nonumber \\
& & \left. - f_{MB/N}(y) \bar s_M(\frac{x}{y}, Q^2) \right ],
\label{eq:s-MCM}
\end{eqnarray}
where $B=\Lambda(\Sigma)$, $M=K(K^{\ast})$, $s_B$ and $\bar s_M$
represent the distributions of strange quark and anti-strange quark
in the baryon and meson, respectively. The $f_{BM/N}(y)$
($f_{MB/N}(y)$) is the fluctuation function which describes the probability
of a nucleon fluctuating into a baryon (meson) with longitudinal momentum fraction $y$.
We have $f_{BM/N}(y) = f_{MB/N} (1-y)$ due to the conservation of momentum and charge,
\begin{eqnarray}\label{ff}
f_{BM/N} (y) =  \sum_{\lambda \lambda^\prime} \int^\infty_0 d
k_\perp^2 \phi^{\lambda \lambda^\prime}_{BM}(y, k_\perp^2)
\phi^{*\,\lambda \lambda^\prime}_{BM}(y, k_\perp^2).
\end{eqnarray}

The baryon-meson wave function, $\phi^{\lambda
\lambda^\prime}_{BM}(y,k_\perp^2)$, and thereby the fluctuation functions $f_{BM}(y)$,
can be calculated using effective meson-baryon-nucleon interaction Lagrangians
\cite{Holtmann:1996be},
\begin{eqnarray}
{\cal{L}}_{NBP} &=& i g_{NBP}\bar N\gamma_5 P B, \\
{\cal{L}}_{NBV} &=& g_{NBV}\bar N\gamma_{\mu} V^{\mu} B  +f_{NBV}\bar N\sigma_{\mu\nu}B(\partial^{\mu} V^{\nu}-\partial^{\nu} V^{\mu}),
\end{eqnarray}
where $g_{NBP}$ and $g_{NBV}$ are the coupling constants. $N$ and
$B$ are spin $1/2$ fields, $P$ and $V$ are pseudoscalar and vector
fields, respectively. The anti-symmetric tensor $\sigma_{\mu\nu}$
is defined as $\sigma_{\mu\nu}=i[\gamma_{\mu},\gamma_{\nu}]/2$.
For the latest calculations for the asymmetry $(s - \bar s)$ using the MCM and more calculation details we refer
the reader to \cite{Cao:2003ny}.

\subsection{Perturbative Contributions}

The perturbative contributions to the strange and anti-strange quarks in the nucleon arising from gluons splitting into quark-antiquark pairs
can be taken into account using perturbative evolution in QCD.
Up to next-to-leading order (NLO) in $\alpha_s$ the probability of a splitting $q \rightarrow q^\prime$ is identical to that of $q \rightarrow {\bar q}^\prime$.
A difference between the probabilities for these splittings arises at NNLO,
which consequently leads to a strange-anti-strange asymmetry \cite{Catani:2004nc}.

The splitting function $P_{ab}$ describing the splitting $b \rightarrow a$ is expanded in terms of $\alpha_s$,
\begin{eqnarray}\label{sfpe}
P_{ab}=\sum_{n=1} \bigg(\frac{\alpha_s}{4\pi}\bigg)^n P_{ab}^{(n-1)}.
\end{eqnarray}
The splitting functions have been calculated up to order $n=3$ \cite{Moch:2004pa,Catani:1994sq}.
Under the assumption of charge conjugation invariance and flavor
symmetry, the splitting functions $P_{ab}$ can be written as \cite{Moch:2004pa,Catani:1994sq},
\begin{eqnarray}\label{generalsplittingfunction}
P_{q_iq_k}&=&P_{\bar q_i\bar q_k}=\delta_{ik}P^V_{qq}+P^S_{qq}\, ,\\
P_{q_i\bar q_k}&=&P_{\bar q_iq_k}=\delta_{ik}P^V_{q\bar
q}+P^S_{q\bar q}\, .
\end{eqnarray}
where the functions $P^S_{qq}$ and $P^S_{q \bar q}$ describe splittings in which the flavor of the quark changes.
The relation $P^S_{qq} = P^S_{q \bar q}$ holds up to NLO in $\alpha_s$.

Only the evolution of flavor nonsinglet (NS) combinations of parton distributions needs to be considered in order to evaluate
$(s - \bar s)$ since it is a flavor NS quantity. The three NS distributions considered in \cite{Catani:2004nc} are
\begin{eqnarray}\label{nsdensity1}
f^{V}&=&\sum_{j=1}^{n_f}(f_{q_j} - f_{\bar q_j}),  \\
f^{\pm}_{q_i}&=&f_{q_i}\pm f_{\bar
q_i}-\frac{1}{n_f}\sum_{j=1}^{n_f}(f_{q_j}\pm f_{\bar q_j}),
\end{eqnarray}
where $n_f$ is the number of flavors.

The evolution equations for these NS distributions are
\begin{eqnarray}\label{ee}
\frac{d\ f^j (x,Q^2)}{d\ln Q^2} = \int_x^1 \,\frac{dz}{z}\,
P^j \left[\frac{x}{z}, \alpha_s(Q^2)\right] \,f^j (z,Q^2)\,
\end{eqnarray}
with $j=\pm,V$.
 The evolution kernels appearing in Eq.~(\ref{ee}) are given by
\begin{eqnarray}
P^{\pm}&=&P^V_{qq} \pm P^V_{q\bar q}, \nonumber \\
P^{V}&=&P^V_{qq} - P^V_{q\bar q}+n_f(P^S_{qq} - P^S_{q\bar q}) \nonumber \\
&\equiv&P^{-}_{ns}+\bigg(\frac{\alpha_s}{4\pi}\bigg)^3P^{(2)S}_{ns},
\end{eqnarray}
where $P^{(2)S}_{ns}$ is the three-loop non-singlet splitting function.

It is convenient to solve the evolution equations in Mellin space where the moments of the distributions $f^j(Q^2)$
and the splitting functions $P^j(x)$,
\begin{eqnarray}\label{mellin transformation}
f^j_N(Q^2) &\equiv& \int_0^1 \,dx \, x^{N-1}\,f^j(x,Q^2),  \\
P^j_N &\equiv& \int_0^1 \,dx \, x^{N-1}\,P^j(x),
\end{eqnarray}
are used.

The Mellin moments of the distribution $ (s- \bar s)$ were found  \cite{Catani:2004nc} to evolve according to,
\begin{eqnarray}\label{pds}
(s-\bar{s})_N(Q^2) &=& U^{-}(Q,Q_0) \,(s-\bar{s})_N(Q_0^2) \nonumber\\
&&-\frac{P_{ns,N}^{(2)S}}{2 \beta_0 N_f}\bigg\{\left[\frac{\alpha_s(Q^2)}{4\pi}\right]^2-\left[\frac{\alpha_s(Q^2_0)}{4\pi}\right]^2\bigg\} \nonumber \\
& & \times \, \left( u^{V}+d^{V} \right) _N(Q^2),
\end{eqnarray}
where the notations $f_s=s$ and $f_{\bar s}=\bar s$ have been used, and $u^{V}$ and $d^{V}$ represent valence $u$-quark and $d$-quark
distributions respectively.

The first term in Eq.~(\ref{pds}) represents nonperturbative contributions to the asymmetry while the second term represents the perturbative contributions.
The evolution operator
\begin{eqnarray}
U^{-}(Q,Q_0)=\exp\bigg\{\int^{  Q^2}_{Q^2_0}\frac{d q^2}{q^2}P^{-}_{ns}[\alpha_s(q^2)]\bigg\}\, ,
\end{eqnarray}
evolves the nonperturbative asymmetry at $Q_0^2$ to higher values of $Q^2$.

In Ref.~\cite{Catani:2004nc} the assumption $(s-\bar{s})_N(Q_0^2)=0$ was adopted and thus only the perturbative contributions to the asymmetry were studied.
The predictions for the asymmetry in the $x$-space, $(s - \bar s)(x,Q^2)$ were obtained via a numerical Mellin inversion.

\section{Total Strangeness Asymmetry}

In order to evaluate the asymmetry in the whole region of $x$ we need to take into account both perturbative and nonperturbative contributions.
Rather than using a numerical Mellin inversion, we apply the inverse Mellin transformation to Eq.~(\ref{pds}) to obtain an expression in $x$ space for
the total asymmetry,
\begin{eqnarray}\label{xspace}
(s-\bar{s})^{\rm Tot} (x,Q^2) = (s-\bar{s})^{\rm NP} (x,Q^2)   + (s-\bar{s})^{\rm P} (x,Q^2),
\end{eqnarray}
with
\begin{eqnarray}\label{eq:pdsxPert}
(s-\bar{s})^{\rm P} (x,Q^2) &=& -\frac{1}{2 \beta_0 N_f}\bigg\{\left[\frac{\alpha_s(Q^2)}{4\pi}\right]^2-\left[\frac{\alpha_s(Q^2_0)}{4\pi}\right]^2\bigg\} \nonumber \\
& & \times \int_x^1 \,\frac{dz}{z}\, P_{ns}^{(2)S}\left(\frac{x}{z} \right)
\left( u^V+d^V \right)(z,Q^2)\,. \nonumber \\
\end{eqnarray}
Both the exact expression and a parameterization form for $P_{ns}^{(2)S}(x)$ are given in \cite{Moch:2004pa}.
The parameterization form, which deviates from the exact expression by less than one part in thousand, is sufficiently accurate for
our calculations,
\begin{eqnarray}
P_{ns}^{(2)S}(x)&\cong& n_f \left\{ [L_0(-163.9x^{-1}-7.208x)+151.49 \right. \nonumber \\
& &+44.51x-43.12x^2+4.82x^3][1-x] \nonumber\\
& & +L_0L_1[-173.1+46.18L_0]+178.04L_0 \nonumber \\
& & \left. +6.892L^2_0+40/27[L^4_0-2L^3_0] \right\}\,,
\end{eqnarray}
with $L_0=\ln x$ and $L_1=\ln (1-x)$.

We use the result obtained in the MCM (Eq.~(\ref{eq:s-MCM}))  to estimate the nonperturbative contributions, $(s-\bar{s})^{\rm NP} (x,Q^2)$.

\section{\label{numerical results}Numerical Results and Discussion}

In the numerical calculations for the nonperturbative contributions we considered Fock states $\Lambda K$, $\Lambda K^*$, $\Sigma K$,
and $\Sigma K^*$, and used the parameters given in \cite{Cao:2003ny}.
The strange and anti-strange quark distributions in the baryons and mesons were calculated using the MIT bag model developed
by the Adelaide group \cite{Signal:1989yc,Schreiber:1991tc,Boros:1999tb} and Massey group \cite{Cao:2001nu}.
The distributions were evaluated at an initial scale of $\mu_0^2=0.23$ GeV$^2$ and evolved to $Q^2=16$ GeV$^2$
using the program given in \cite{MiyamaK96}.

\begin{figure}
\begin{minipage}{\columnwidth}
\centering
\includegraphics[width=2.9in]{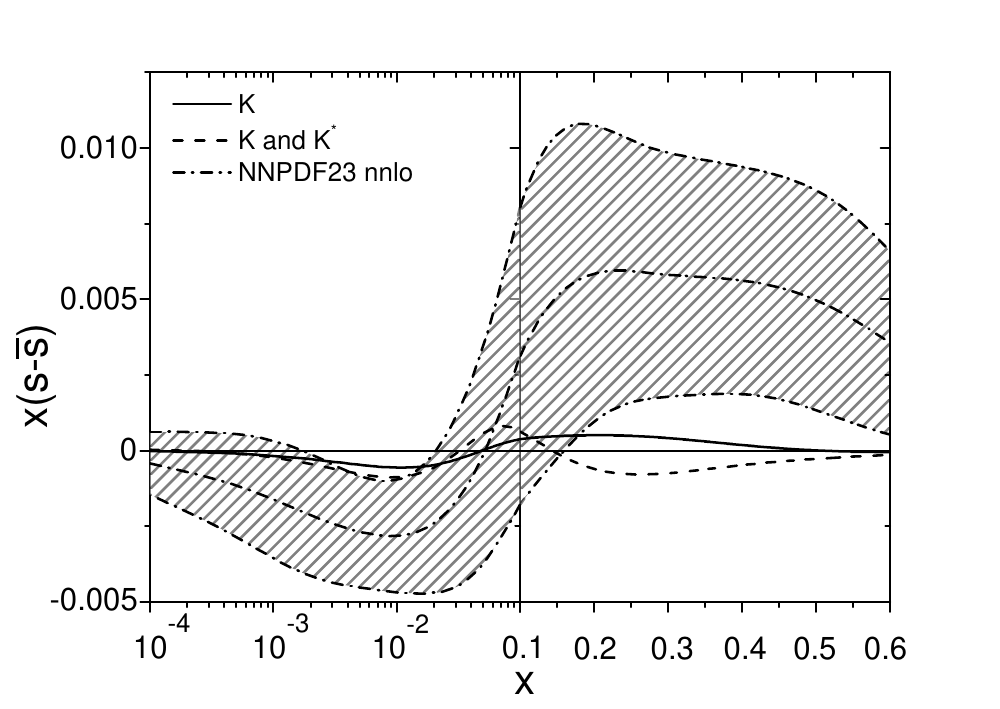}
\end{minipage}
\caption{\small Nonperturbative contributions to the asymmetry $x(s-\bar{s})$ at $Q^2=16$ GeV$^2$ calculated with the meson cloud model, in comparison with
the total asymmetry obtained with the NNPDF2.3 NNLO PDF set.
The solid and dashed curves are the MCM results considering $K$-meson contributions and  $K$- and $K^*$-mesons contributions, respectively.
The shaded area represents one-sigma uncertainty band for the NNPDF2.3 set.
\label{fig_np}}
\end{figure}

The results, in comparison with the asymmetry obtained with the NNPDF2.3 PDF set \cite{NNPDF2.3}, are shown in Fig.~\ref{fig_np}.
One can see that the non-perturbation calculations for the asymmetry are generally much smaller than the total asymmetry obtained with the NNPDF2.3 PDF set.

The perturbative contributions depend on the initial scale chosen (see Eq.~(\ref{eq:pdsxPert})).
Two values for the initial scale, $Q_0=0.51$ GeV and $Q_0=1.1$, were used in our numerical evaluations as in \cite{Catani:2004nc}.
It is worth noting that the calculations depend on the up and down valence distributions at high $Q^2$ where hard scattering data available, rather than the distributions at the initial scale $Q_0^2$.
The valance distributions of the nucleon are reasonably well determined via global fits. Although the $d^V$ and $u^V$ obtained with different PDF sets may not agree within the quoted uncertainties,
the differences are generally smaller than $10\%$. Thus we can expect similar level of uncertainties due to the choice of PDF sets
in the calculations for the perturbative contributions to the strangeness asymmetry using Eq.~(\ref{eq:pdsxPert}).
We employed the ABM11 NNLO PDF set for the $d^V$ and $u^V$ at $Q^2=16$ GeV$^2$ in our evaluations.

\begin{figure}
\begin{minipage}{\columnwidth}
\centering
\includegraphics[width=2.9in]{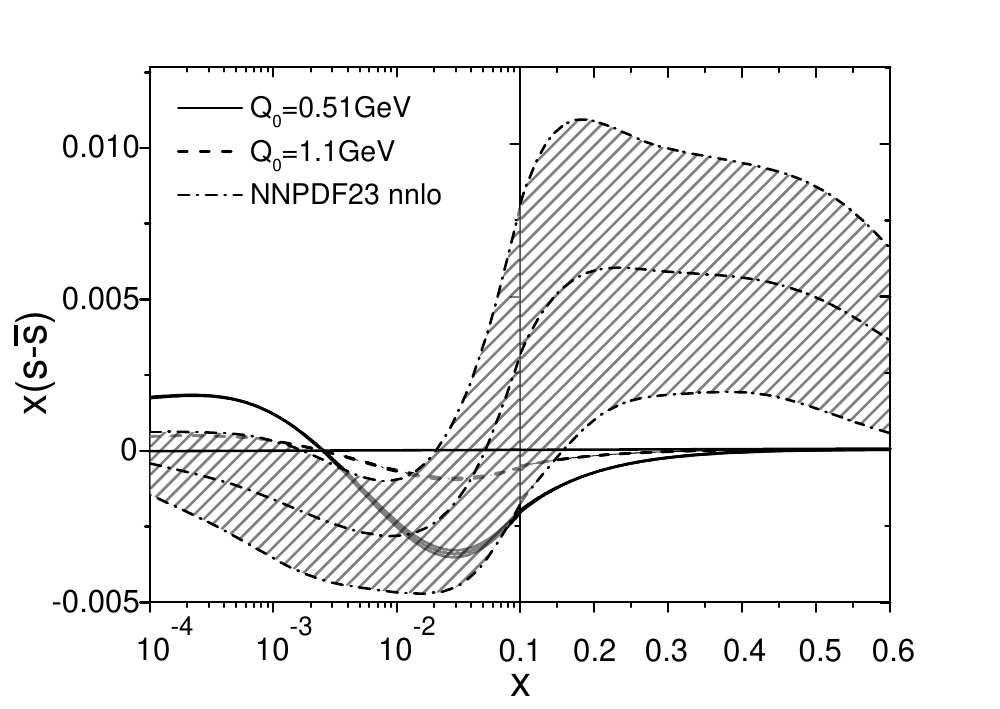}
\end{minipage}
\caption{\small Perturbative contributions to the asymmetry $x(s-\bar{s})$ at $Q^2=16$ GeV$^2$,
in comparison with the total asymmetry obtained with the NNPDF2.3 NNLO PDF set.
The solid and dashed curves are the results calculated with $Q_0=0.51$ GeV and $Q_0=1.1$ GeV, respectively.
The bands in the perturbative calculations represent the one-sigma range due to the uncertainties associated with the $d^V$ and $u^V$ of the ABM11 PDF set.
The shaded area represents one-sigma uncertainty band for the NNPDF2.3 set.
\label{fig_pert}}
\end{figure}

The perturbative contributions to the asymmetry $x(s-\bar s)$, in comparison with the asymmetry obtained with the NNPDF2.3 PDF set \cite{NNPDF2.3}, are shown in Fig.~\ref{fig_pert}.
The results are in agreement with those presented in \cite{Catani:2004nc}.
Note that we have used an inverse Mellin transformation of Eq.~(\ref{pds}) to obtain the expression for perturbative contributions
in the $x$ space (Eq.~(\ref{eq:pdsxPert})) while the Mellin inversion was done numerically in \cite{Catani:2004nc}.
The perturbative contributions in the small $x$-region ($x< 0.05$) tend to agree with results obtain with the NNPDF2.3 NNLO PDF set, which suggests that the perturbative contributions are
the dominant mechanism for the generation of this asymmetry at small $x$.

From Figs.~\ref{fig_np} and \ref{fig_pert} one can see that the nonperturbative contributions dominate in the region $x>0.1$ while the perturbative contributions dominate in the region $x<0.1$.
The nonperturbative contributions are smaller in magnitude than the perturbative contributions, which indicates that the asymmetry may be most easily measured in the region of $x$ around 0.03.

\begin{figure}
\begin{minipage}{\columnwidth}
\centering
\includegraphics[width=2.9in]{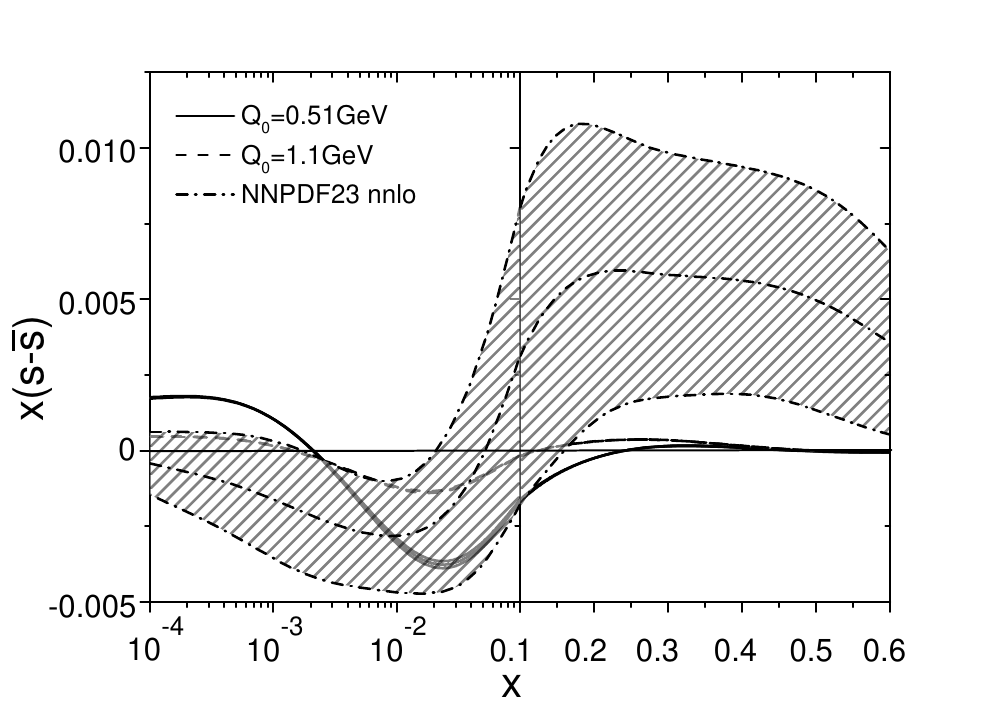}
\end{minipage}
\caption{\small The total asymmetry $x(s-\bar{s})$ at $Q^2=16$ GeV$^2$
including both the perturbative contributions and nonperturbative contributions which are calculated with only $K$ mesonic cloud.
The solid and dashed curves are the results obtained with $Q_0=0.51$ GeV and $Q_0=1.1$ GeV in the perturbative calculations.
The bands represent the one-sigma range due to the uncertainties associated with the $d^V$ and $u^V$ at $Q^2=16$ GeV$^2$.
The shaded area represents one-sigma uncertainty band for the NNPDF2.3 set.
\label{fig_totK}}
\end{figure}

\begin{figure}
\begin{minipage}{\columnwidth}
\centering
\includegraphics[width=3.5in]{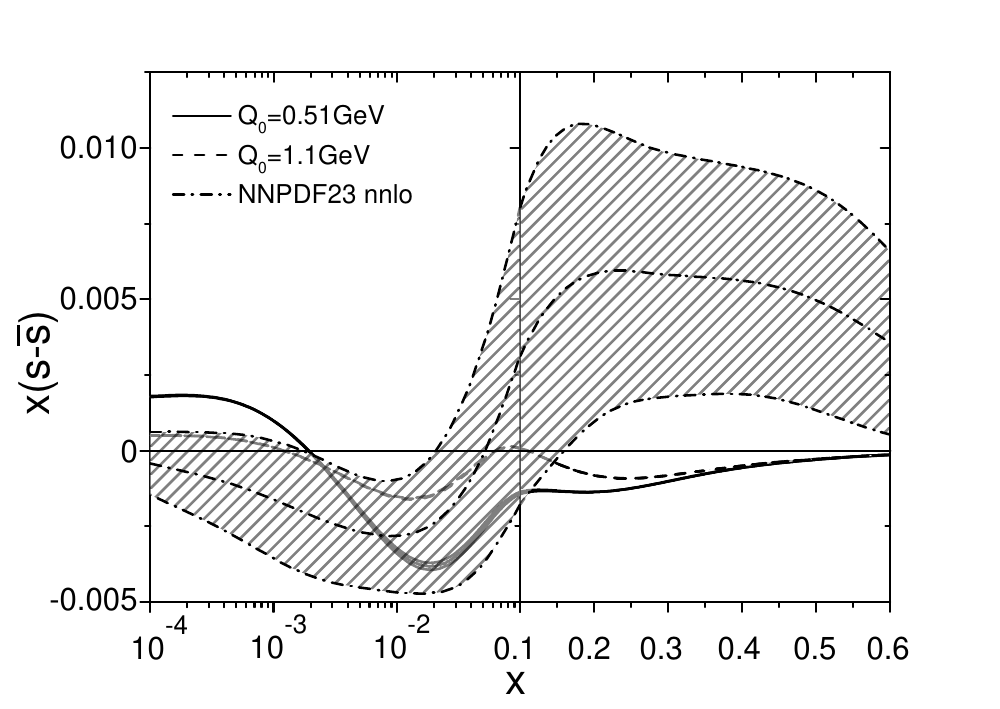}
\end{minipage}
\caption{\small Same as Fig.~\ref{fig_totK} but both $K$ and $K^*$ mesonic clouds are considered in the nonperturbative contributions.
\label{fig_totKK*}}
\end{figure}

The calculations for the total asymmetry including both perturbative and nonperturbative contributions are shown in Figs.~\ref{fig_totK} and \ref{fig_totKK*}.
Only the $K$-meson contributions were considered in the nonperturbative calculations for Fig.~\ref{fig_totK}, while both
$K$- and $K^*$-mesons contributions were included in the nonperturbative calculations for Fig.~\ref{fig_totKK*}.

Figs.~\ref{fig_totK} and \ref{fig_totKK*} suggest that the asymmetry may have more than one node, in contrast to the
assumption of only one node for the asymmetry commonly used in most global analyses of the PDFs \cite{Martin:2009iq,Lai:2007dq,BaronePZ00,Barone:2006xj}.
In the case of one node, the calculated asymmetry presented in this work can be described using the
parameterization suggested in \cite{Lai:2007dq},
\begin{eqnarray}
s_{-}(x, Q) = A_0 x^{A_1} (1-x)^{A_2}  {\rm tan}^{-1} \left[ c x^a \left(1 - \frac{x}{b} \right) e^{d x + e x^2} \right], \label{eq:s-CTEQ6.5S}
\end{eqnarray}
where $A_0$, $A_1$, $A_2$, $a$, $b$ $c$, $d$, and $e$ are parameters.
It is very difficult to describe our calculated asymmetry using other forms simpler than Eq.~(\ref{eq:s-CTEQ6.5S}) and involving fewer parameters.
It is interesting to note that when only $K$-mesons are considered in the nonperturbative contributions the calculated asymmetry exhibits a
shape similar to that determined with the NNPDF2.3 NNLO PDF set (see Fig.~\ref{fig_totK}), though the magnitudes at the region of $x>0.2$ are rather different.

The current experimental constraints for the strangeness of the nucleon come mostly from the neutrino dimuon production data, and are not
strong enough to allow
a meaningful fit for the strangeness distributions using the many numbers of parameters of Eq.~(\ref{eq:s-CTEQ6.5S}) \cite{Lai:2007dq}.
This highlights the need for more accurate future experiments to constrain the strange and anti-strange distributions of the nucleon.
The Large Hadron Collider has the great potential to provide new information for the strange content of the nucleon, as shown by a recent study
by the ATLAS Collaboration \cite{ATLAS}.

\begin{table*}
\caption{$\langle x(s-\bar{s})\rangle$ at $Q^2=16$ GeV$^2$}
\label{comparison}
\begin{tabular*}{\textwidth}{@{\extracolsep{\fill}}lrrrl@{}}
\hline
$Q_0$(GeV)& \multicolumn{1}{c}{P} & \multicolumn{1}{c}{P+NP without $K^{\ast}$} & \multicolumn{1}{c}{P+NP including $K^{\ast}~~$}\\
\hline
0.51 & $-5.39\times10^{-4}$ & $-4.09\times10^{-4}$ & $-7.09\times10^{-4}$ \\
\hline
1.10 & $-1.27\times10^{-4}$ & $3.33\times10^{-6}$ & $-2.96\times10^{-4}$ \\
\hline
\end{tabular*}
\end{table*}
To investigate the possible effects on the NuTeV measurement of the Weinberg angle we
calculated the second moment of the asymmetry $\langle x(s - \bar s) \rangle  = \int_0^1 d x x( s-\bar s)$.
The results are given in Table \ref{comparison} where  P and NP stand for the perturbative and nonperturbative contributions.
The results depend on the model parameters -- the initial scale chosen in the perturbative calculations and whether the $K^*$ mesonic cloud is included
in the calculations for the nonperturbative contributions.
A value of $\langle x(s - \bar s) \rangle \sim 0.004$ would be required if the anomalous NuTeV result was caused only by the strange quark asymmetry.
Our calculations suggest that the second moment $\langle x(s - \bar s) \rangle$ is negative, but is about an order of magnitude too small to have any
significant effect on the NuTeV result for the Weinberg angle.
Thus other mechanisms, as discussed in \cite{Davidson:2001ji,Londergan:2003ij,Kulagin:2003wz,Kretzer:2003wy},
could be responsible for the NuTeV anomaly.

\section{\label{summary}Summary}
In this paper, we analyzed the strange and anti-strange quark asymmetry in the nucleon sea.
We have taken into account  both the nonperturbative contributions as well as the contributions from perturbative QCD evolution.
The nonperturbative contributions were calculated using the meson cloud model in which
the baryon-meson Fock state, $N\rightarrow \Lambda K(K^{\ast})$, and $N\rightarrow \Sigma K(K^{\ast})$ were included.
The perturbative contributions arises at the NNLO in perturbative QCD because the splitting function for the splitting $q \rightarrow q^\prime$
differs from that for the splitting $q \rightarrow {\bar q}^\prime$ starting from that order.

The nonperturbative contributions are dominant in the region $x \gtrsim 0.1$ while the perturbative contributions are dominant in the smaller $x$ region.
It was found that the perturbative contributions are generally a few times larger in magnitude than
the nonperturbative contributions, which suggests that the best region to detect this asymmetry experimentally is in the region of $x$ around 0.03.
We found that the asymmetry may have more than one node which could shed light on the parameterization of the
strange and anti-strange quark distributions in the global analysis of the parton distribution functions.
More experiments directly detecting the strangeness of the nucleon are highly desired.
We found that the asymmetry is not large enough to have any significant effects on the NuTeV measurement for the Weinberg angle.

\begin{acknowledgements}
F.-G. thanks D. Sutherland for discussion on the Mellin transformation. G.-Q. Feng is very grateful to Professor B.-S. Zou for valuable discussions.
This work is supported by the National Natural Science Foundation of China (Project Nos. 11035006, 10675022, 10975018, and 11175020),
the Chinese Academy of Sciences (Project No. KJCX2-EW-N01), and the Fundamental Research Funds for the Central Universities in China.
\end{acknowledgements}
\bibliographystyle{unsrt}
\bibliography{smsbarbibli}

\begin{thebibliography}{10}

\bibitem{PDF_MRST00}
A.~D. Martin, R.~G. Roberts, W.~J. Stirling, and R.~S. Thorne.
\newblock {\em Eur. Phys. J.}, C14:133, 2000.

\bibitem{PDF_CTEQ5}
H.-L. Lai et~al.
\newblock {\em Eur. Phys. J.}, C12:375, 2000.

\bibitem{PDF_GRV98}
M.~Gluck, E.~Reya, and A.~Vogt.
\newblock {\em Eur. Phys. J.}, C5:461, 1998.

\bibitem{PDF_Alekhin01}
S.~I. Alekhin.
\newblock {\em Phys. Rev.}, D63:094022, 2001.

\bibitem{KuhnCL09}
S.~E. Kuhn, C.~P. Chen, and E.~Leader.
\newblock {\em Prog. Part. Nucl. Phys.}, 73:1, 2009.

\bibitem{BurkardtMN10}
M.~Burkardt, C.~A. Miller, and W.~D. Nowak.
\newblock {\em Rep. Prog. Phys.}, 73:016201, 2010.

\bibitem{Kusina:2012vh}
A.~Kusina, T.~Stavreva, S.~Berge, F.I. Olness, I.~Schienbein, et~al.
\newblock {\em hep-ph/1203.1290}, 2012.

\bibitem{Signal:1987gz}
A.~I. Signal and Anthony~William Thomas.
\newblock {\em Phys. Lett.}, B191:205, 1987.

\bibitem{Davidson:2001ji}
S.~Davidson, S.~Forte, P.~Gambino, N.~Rius, and A.~Strumia.
\newblock {\em JHEP}, 0202:037, 2002.

\bibitem{Cao:2003ny}
F.-G. Cao and A.~I. Signal.
\newblock {\em Phys. Lett.}, B559:229--234, 2003.

\bibitem{Zeller:2001hh}
G.~P. Zeller et~al.
\newblock {\em Phys. Rev. Lett.}, 88:091802, 2002.

\bibitem{PDG2012}
J.~Beringer et~al.
\newblock {\em Phys. Rev.}, D86:010001, 2012.

\bibitem{Martin:2009iq}
A.~D. Martin, W.~J. Stirling, R.~S. Thorne, and G.~Watt.
\newblock {\em Eur. Phys. J.}, C63:189--285, 2009.

\bibitem{Lai:2007dq}
H.~L. Lai et~al.
\newblock {\em JHEP}, 04:089, 2007.

\bibitem{Lai:2010}
H.~L. Lai et~al.
\newblock {\em Phys. Rev.}, D82:074024, 2010.

\bibitem{JR09}
P.~Jimenez-Delgado and E.~Reya.
\newblock {\em Phys. Rev.}, D79:074023, 2009.

\bibitem{ABKM10}
S.~Alekhin, J.~Blumlein, S.~Klein, and S.~Moch.
\newblock {\em Phys. Rev.}, D81:014032, 2010.

\bibitem{ABM11}
S.~Alekhin, J.~Blumlein, and S.~Moch.
\newblock {\em Phys. Lett.}, B697:127, 2011.

\bibitem{BaronePZ00}
V.~Barone, C.~Pascaud, and F.~Zomer.
\newblock {\em Eur. Phys. J.}, C12:2, 2000.

\bibitem{Barone:2006xj}
V.~Barone, C.~Pascaud, B.~Portheault, and F.~Zomer.
\newblock {\em JHEP}, 01:006, 2006.

\bibitem{NNPDF1.2}
R.D. Ball et~al.
\newblock {\em Nucl. Phys.}, B823:195, 2009.

\bibitem{NNPDF2.1}
R.D. Ball et~al.
\newblock {\em Nucl. Phys.}, B849:296, 2011.

\bibitem{NNPDF2.3}
R.D. Ball et~al.
\newblock {\em arXiv:1207.1303[hep-ph]}.

\bibitem{J2010}
P~Jimenez-Delgado.
\newblock {\em Phys. Lett.}, B689:177, 2010.

\bibitem{AKP09}
S.~Alekhin, S.~Kulagin, and R.~Petti.
\newblock {\em Phys. Lett.}, B675:433, 2009.

\bibitem{Cao:2011qh}
Fu-Guang Cao.
\newblock {\em AIP Conf. Proc.}, 1418:91--98, 2011.

\bibitem{Thomas:1983fh}
Anthony~William Thomas.
\newblock {\em Phys. Lett.}, B126:97, 1983.

\bibitem{Burkardt:1991di}
M.~Burkardt and Brian Warr.
\newblock {\em Phys. Rev.}, D45:958--964, 1992.

\bibitem{Brodsky:1996hc}
Stanley~J. Brodsky and Bo-Qiang Ma.
\newblock {\em Phys. Lett.}, B381:317--324, 1996.

\bibitem{Holtmann:1996be}
H.~Holtmann, A.~Szczurek, and J.~Speth.
\newblock {\em Nucl. Phys.}, A596:631--669, 1996.

\bibitem{Cao:1999da}
Fu-Guang Cao and A.~I. Signal.
\newblock {\em Phys. Rev.}, D60:074021, 1999.

\bibitem{Melnitchouk:1999mv}
W.~Melnitchouk and M.~Malheiro.
\newblock {\em Phys. Lett.}, B451:224--232, 1999.

\bibitem{Signal:2000xy}
A.~I. Signal and F.-G. Cao.
\newblock {\em Nucl. Phys.}, A680:43--47, 2000.

\bibitem{Christiansen:1998dz}
H.~R. Christiansen and J.~Magnin.
\newblock {\em Phys. Lett.}, B445:8--13, 1998.

\bibitem{Cao:2004pr}
F.-G. Cao and A.~I. Signal.
\newblock {\em Nucl. Phys. Proc. Suppl.}, 128:30--36, 2004.

\bibitem{Alwall:2004rd}
Johan Alwall and Gunnar Ingelman.
\newblock {\em Phys. Rev.}, D70:111505, 2004.

\bibitem{Wakamatsu:2004pd}
M.~Wakamatsu.
\newblock {\em Phys. Rev.}, D71:057504, 2005.

\bibitem{Alwall:2005xd}
Johan Alwall and Gunnar Ingelman.
\newblock {\em Phys. Rev.}, D71:094015, 2005.

\bibitem{Wei:2007nb}
F.~X. Wei and B.~S. Zou.
\newblock {\em Phys. Lett.}, B660:501--504, 2008.

\bibitem{Diehl:2007uc}
M.~Diehl, Th. Feldmann, and P.~Kroll.
\newblock {\em Phys. Rev.}, D77:033006, 2008.

\bibitem{Traini:2011tc}
Marco Traini.
\newblock {\em Phys. Lett.}, B707:523--528, 2012.

\bibitem{Catani:2004nc}
Stefano Catani, Daniel de~Florian, German Rodrigo, and Werner Vogelsang.
\newblock {\em Phys. Rev. Lett.}, 93:152003, 2004.

\bibitem{Moch:2004pa}
S.~Moch, J.~A.~M. Vermaseren, and A.~Vogt.
\newblock {\em Nucl. Phys.}, B688:101--134, 2004.

\bibitem{Catani:1994sq}
S.~Catani and F.~Hautmann.
\newblock {\em Nucl. Phys.}, B427:475--524, 1994.

\bibitem{Signal:1989yc}
A.~I. Signal and Anthony~William Thomas.
\newblock {\em Phys. Rev.}, D40:2832--2843, 1989.

\bibitem{Schreiber:1991tc}
Andreas~W. Schreiber, A.~I. Signal, and Anthony~William Thomas.
\newblock {\em Phys. Rev.}, D44:2653--2662, 1991.

\bibitem{Boros:1999tb}
C.~Boros and Anthony~William Thomas.
\newblock {\em Phys. Rev.}, D60:074017, 1999.

\bibitem{Cao:2001nu}
Fu-Guang Cao and A.I. Signal.
\newblock {\em Eur. Phys. J.}, C21:105--114, 2001.

\bibitem{MiyamaK96}
M.~Miyama and S.~Kumano.
\newblock {\em Comput. Phys. Commun.}, 94:185, 1996.

\bibitem{ATLAS}
G.~Aad et~al.
\newblock {\em Phys. Rev. Lett.}, 109:012001, 2012.

\bibitem{Londergan:2003ij}
J.~T. Londergan and Anthony~William Thomas.
\newblock {\em Phys. Rev.}, D67:111901, 2003.

\bibitem{Kulagin:2003wz}
Sergey~A. Kulagin.
\newblock {\em Phys. Rev.}, D67:091301, 2003.

\bibitem{Kretzer:2003wy}
Stefan Kretzer et~al.
\newblock {\em Phys. Rev. Lett.}, 93:041802, 2004.

\end{thebibliography}
\end{document}